\documentclass[runningheads]{IEEEtran}
\usepackage{cite}
\usepackage{amsmath,amssymb,amsfonts}
\usepackage{algorithmic}
\usepackage{graphicx}
\usepackage{textcomp}
\usepackage{xcolor, colortbl}
\usepackage{wrapfig}
\usepackage{color,soul}
\usepackage{multirow}
\usepackage{url}
\usepackage{tikz}

\title{
An Evolutionary Algorithm for Task Scheduling in Crowdsourced Software Development}

\author{Razieh Saremi\inst{1} \and
Hardik Yagnik\inst{1}\and
Julian Togelius\inst{2}\and
Ye Yang\inst{1}\and Guenther Ruhe\inst{3}}
\authorrunning{R. Saremi et al.}
\institute{Stevens Institute of Technology, Hoboken NJ, USA \and 
New York University, NYC NY, USA \and University of Calgary, Calgary, Alberta, Canada \\
\email{\{rsaremi,hyagnik1,  yyang4\}@stevens.edu,  julian.togelius@nyu.edu, ruhe@ucalgary.ca}\\
}

\begin{document}

\maketitle
\small
\begin{abstract}

The complexity of software tasks and the uncertainty of crowd developer behaviors make it challenging to plan crowdsourced software development (CSD) projects. In a competitive crowdsourcing marketplace, competition for shared worker resources from multiple simultaneously open tasks adds another layer of uncertainty to potential outcomes of software crowdsourcing. 
These factors lead to the need for supporting CSD managers with automated scheduling to improve the visibility and predictability of crowdsourcing processes and outcomes. 
To that end, this paper proposes an evolutionary algorithm-based task scheduling method for crowdsourced software development.
The proposed evolutionary scheduling method uses a multiobjective genetic algorithm to recommend optimal task start date. The method uses three fitness functions, based on project duration, task similarity, and task failure prediction, respectively. The task failure fitness function uses a neural network to predict the probability of task failure with respect to a specific task start date.  The proposed method then recommends the best tasks' start dates for the project as a whole and each individual task so as to achieve the lowest project failure ratio. Experimental results on 4 projects demonstrate that the proposed method has the potential to reduce project duration by a factor of 33-78\%.


\end{abstract}

\small

\small
\section{Introduction}

Crowdsourced Software Development (CSD) has been increasingly adopted in modern software development practices as a way of leveraging the wisdom of the crowd to complete software mini-tasks faster and cheaper \ cite{stol2014two}\cite{saremi2017leveraging}. 
In order for a CSD platform to function efficiently, it must address both the needs of task providers as demands and crowd workers as suppliers. Mismatches between these needs might lead to task failure in the CSD platform. In general, planning for CSD tasks is challenging \cite{stol2014two}, because software tasks are complex, independent, and require potential task-takers pertaining to specialized skill-sets. The challenges stem from three factors:
1) task requesters typically need to simultaneously monitor and control a large number of mini-tasks decomposed suitable for crowdsourcing; 2) task requesters generally have little to no control over how many and how dedicated workers will engage in their tasks, and 3) task requesters are competing for shared worker resources with other open tasks in the market. 

To address these challenges, existing research has explored various methods and techniques to bridge the information gap between demand and supply in crowdsourcing-based software development. Such research includes studies towards developing a better understanding of worker motivation and preferences in CSD \cite{faradani2011s}\cite{gordon1961general}\cite{difallah2016scheduling}\cite{yang2015award}, studies focusing on predicting task failure \cite{khanfor2017failure}\cite{khazankin2011qos}; studies employing modeling and simulation techniques to optimize CSD task execution processes\cite{saremi2018hybrid}\cite{saremi2019ant}\cite{saremi2021crowdsim}; and studies for recommending the most suitable workers for tasks \cite{yang2016should} and developing methods to create crowdsourced teams\cite{wang2018solving} \cite{yue2015evolutionary}. 
The existing work lacks effective supports for analyzing the impact of task similarity and task arrival date on task failure. In this study, we approach these gaps by proposing a task scheduling method leveraging the combination of evolutionary algorithms and neural networks. 

The objective of this work is to propose a task scheduling recommendation framework to support CSD managers to explore options to improve the success and efficiency of crowdsourced software development. To this end, we first present a motivating example to highlight the practical needs for software crowdsourcing task scheduling. Then we propose a task scheduling architecture based on an evolutionary algorithm. This algorithm seeks to reduce predicted task failure while at the same time shortening project duration and managing the level of task similarity in the platform. The system takes as input a list of tasks, their durations, and their interdependencies. It then finds a plan for the overall project, meaning an assignment of each task to a specific day after the start of the project.
Also, the system checks the similarity among parallel open tasks to make sure the task will attract the most suitable available worker to perform it upon arrival.
The evolutionary task scheduler then recommends a task schedule plan with the lowest task failure probability per task to arrive in the platform. We applied the proposed scheduling framework to 4 CSD projects in our database. The result confirmed that the proposed method has the potential of reducing the project duration between 33-78\%.

The proposed system represents a task scheduling method for competitive crowdsourcing platforms based on the workflow of TopCoder\footnote{ \url{https://www.topcoder.com/}}, one of the primary software crowdsourcing platforms.  The recommended schedule provides improved duration while decreasing task failure probability and number of open similar tasks upon task arrival in the platform. 

The remainder of this paper is structured as follows. Section II introduces a motivating  example that inspires this study. Section III outlines our research design and methodology. Section IV presents the case study and model evaluation.Section V presents a review of related works, and Section VI presents the conclusion and outlines a number of directions for future work.

\small
\section{Motivating Example}

\begin{figure*}[ht!]
\includegraphics[width=1\textwidth]{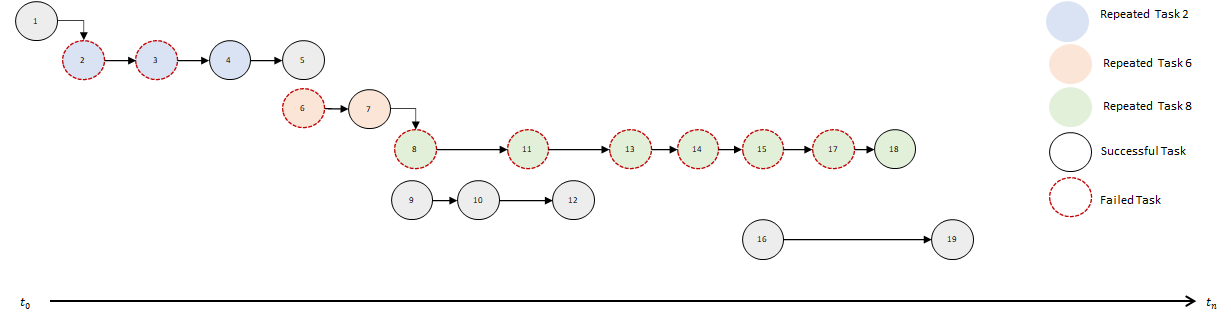}
\caption{Overview of motivating example }
\label{Motivation}
\end{figure*}

The motivating example illustrates a real-world crowdsourcing software development (CSD) project on the TopCoder platform. It consists of 19 tasks with a project duration of 110 days. The project experienced a 57\% task failure ratio, meaning 11 of the 19 tasks failed. More specifically, Task 14 and 15 failed due to client request, and Task 3 failed due to unclear requirements. The remaining eight tasks (i.e. Task 1, 2, 4, 5, 8, 11, 13, and 17) failed due to zero submissions. 
An in-depth analysis revealed that the eight failed tasks due to zero submissions were basically three tasks (i.e. tasks 2, 4, and 8) that were re-posted after each failure to be successfully completed as the new task.

As illustrated in Figure \ref{Motivation}, Task 2 was cancelled and re-posted as Tasks 5 and 7. It was completed as Task 7 with changes in the monetary prize and task type. Task 4 was cancelled and re-posted as Task 6 which was completed with no modification. Task 8 also failed and re-posted six times as Tasks 11, 13, 14, 15, 17, and 18. Each re-posting modified Task 8 in terms of the monetary prize, task type, and required technology. Finally, the task was completed as Task 18. 
Studying task 18 revealed that it arrived with 5 similar tasks with similarity degree of 60\% in the platform. Also task failure probability on the arrival day for task 18 was 17\%.


This observation motivated us to develop the scheduling method presented in this paper. This method can reduce the probability of task failure in the platform, while simultaneously controlling the level of task similarity on task arrival day per project.

\begin{figure*}[ht!]
\centering
\includegraphics[width=1\textwidth,height=0.5\textwidth]{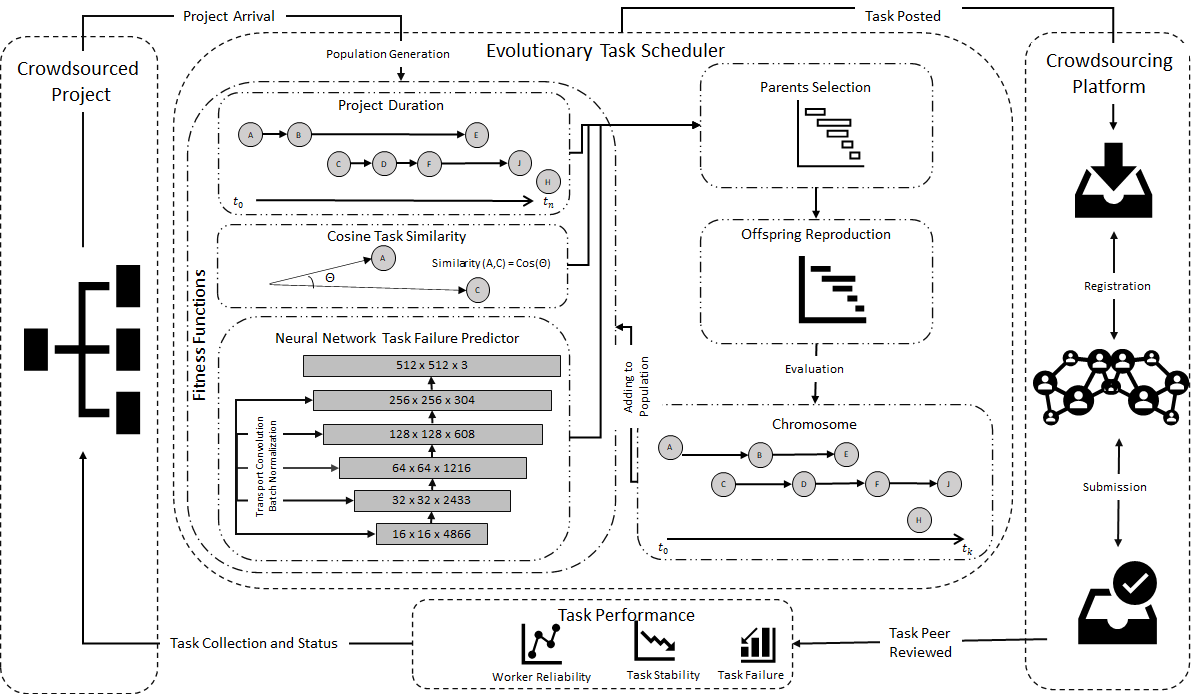}
\caption{Overview of Evolutionary Scheduling Architecture}
\label{diagram}
\end{figure*}

\small

\section{Research Methodology}\label{architechture}

To solve the scheduling problem, we use a genetic algorithm to schedule tasks based on the minimum probability of task failure and degree of similarity among list of parallel tasks in the project. We adapted dominant CSD project attributes from existing studies
\cite{stol2014two}\cite{saremi2017leveraging}\cite{yang2015award}\cite{saremi2015empirical} \cite{mejorado2020study}
and defined metrics used in the scheduling framework.
Then we added the neural network model from \cite{iceis21} to predict the probability of task failure per day. 
In the method presented here, task arrival date is suggested based on the degree of task similarity among parallel tasks in the project and the probability of task failure in the platform based on available tasks and reliability of available workers to make a valid submission in the platform. Figure \ref{diagram} presents the overview of the task scheduling architecture. List of tasks of a crowdsourced project is uploaded in the \textit{evolutionary task scheduler}. The \textit{task dependency} orders tasks based on the project dependency and tasks duration. In parallel, \textit{task similarity} calculates the similarity among tasks to make sure batch of arrival tasks follow the optimum similarity level. Then, the \textit{task failure predictor} analyzes the probability of failure of an arriving task in the platform based on the number of similar tasks available on arriving day, average similarity, task duration, and associated monetary prize. In next step, the \textit{parent selection} selects the most suitable set of parents to use for \textit{reproduction}, i.e. the generation of new chromosomes. In every generation, all chromosomes are evaluated by the fitness functions. 
The result of task performance in the platform is to be collected and reported to the client along with the input used to recommend the posting date.

\subsection{Evolutionary Task Scheduling Method}

We pose crowdsourced task scheduling as a multi-objective optimization problem with dynamic characteristics that requires agility in tracking the changes of task similarity and probability of task failure over time. 
Therefore, we presume that the objectives exhibit partial conflict (this was also borne out by observation) so that it will be unlikely to find a single solution that satisfies all objectives. In other words, a successful optimization finds a Pareto front of non-dominated solutions. To that end, we propose a novel evolutionary task scheduling method which combines multiple objectives. The presented method integrates artificial neural network with the non-dominated sorting genetic algorithm (NSGA-II) \cite{996017}, which is a well-known method for finding sets of solutions for multi-objective problems.

\subsubsection{Task Dependency}
A crowdsourced project (${CP_{K}}$) is a sequence of time dependent tasks that needs to be completed during specific period of time. In this study, the time sequence of an arriving task is considered as a measurement of the series of parallel or sequential tasks that followed finish-to-start task dependency techniques ${TD_{i}}$. Task (n), ${T_{n}}$, is parallel (${Pl_{i}}$) with task (n-1), ${T_{n-1}}$, if it starts before ${T_{n-1}}$ submission deadline. And ${T_{n}}$ is a sequential task (${Sl_{i}}$) for ${T_{n-1}}$ if it starts after ${T_{n}}$ submission deadline. The project duration (${CPD_{K}}$) is defined as the total duration between first task registration start data (${TR_{1}}$) and last task submission end date(${TS_{n}}$). 

\begin{equation}
\scriptsize
{CP_{k}} = {(\{{T_{i}}\}, \{{TD_{i}}\},{CPD_{k}})}
\end{equation}
\[
where
\scriptsize
	\begin{cases}
	    \begin{cases}
            {TD_{i}} = \begin{cases}
                        1 &{Pl_{i}} = 1 \\
                        0 & {Sl_{i}} = 1
                       \end{cases} \\
            \text{i= 1, 2, 3, …, n}
        \end{cases}\\
        \\
        \begin{cases}
            {CPD_{k}} =  {{TS_{n}} - {TR_{1}}}
            \\
           \text{k = 1, 2, 3, …, m} 
        \end{cases}
    \end{cases}
\]
\small 

\subsubsection{Task Similarity}
Task similarity is defined as the degree of similarity between a set of simultaneously open tasks. We calculate the tasks’ local distance from each other to calculate a task similarity factor, based on different task features  such as detailed task requirements, task monetary prize, task type, task  registration and submissions date, required technology and platform. 

\textit{Def.1}: Task local distance (${Dis_{i}}$) is a tuple of tasks’ attributes in the data set. In respect to introduce variables in table \ref{Sim}, task local distance is:

\begin{equation}
{Dis_{i}} = {({Prize_{i}}, {TR_{i}}, {TS_{i}}, {Type_{i}}, {Techs_{i}}, {PLTs_{i}}, {Req_{i}})}
\end{equation}

\textit{Def.2}: Task Similarity ($ {Sim_{i,j}} $) between two tasks $ {T_{i}} $ and $ {T_{j}} $ is defined as the is the dot product and magnitude of the local distance of two tasks:

\begin{equation}
{Sim_{i,j}} = \frac{\sum_{i,j=0}^{n}{Dis_{(i,j)}}} 
{\sum_{i=0}^{n}\sqrt{{Dis_{i}}}*\sum_{j=0}^{n}\sqrt{{Dis_{j}}}} 
\end{equation}


\begin{small}
\begin{table}[!ht]
\scriptsize
\caption{Features used to measure task distance} 
\centering 
\begin{tabular}{p{3.5cm} p{6cm}}
\hline
Feature & Description of distance measure $ {Dist_{i}} $ \\ 
\hline
Distance of task monetary prize ($\Delta$MP) & (${Prize_{i}}$ - ${Prize_{j}}$ ) = ${Prize_{Max}}$ \\
Distance of task registration start date ($\Delta$TR)	& (${TR_{i}}$ - ${TR_{j}}$) = ${DiffTR_{Max}}$ \\
Distance of task submission end date ($\Delta$TS)	& (${TS_{i}}$ - ${TS_{j}}$) = ${Diff_TS_{Max}}$ \\
Task Type & (${Type_{i}}$ == ${Type_{j}}$) ? 1 : 0 \\
Technology (Techs) & Match(${Tech_{i}}$:${Tech_{j}}$)=${NumberofTechs_{Max}}$ \\ 
Platform (PLTs) & (${PL_{i}}$ == ${PL_{j}}$) ? 1 : 0 \\
Textual Task Requirement & (${Req_{i}} * {Req_{j}})/(|{Req_{i}}|*|{Req_{i}}|$) \\

\hline
\label{Sim}
\end{tabular}
\end{table}
\end{small}

\subsubsection{Task Failure Predictor}

To predict probability of task failure, a fully connected feed forward neural network was trained \cite{iceis21}.
It is reported that task monetary prize and task duration \cite{yang2015award}\cite{faradani2011s} are the most important factors in raising competition level for a task. In this research, we are adding the variables considered in our observations (i.e number of open tasks and average task similarity in the platform) to the reported list of important factors as input features to train the neural network model.
The network is configured with five layers of size 32, 16, 8, 4, 2, and 1. We applied  a  K-fold (K=10)  cross-validation method  on  the train/test group to train  the prediction of task  failure probability in the neural network model. Also, We used early stopping to avoid over-fitting. The trained model provided a loss of 0.04 with standard deviation of 0.002. Interestingly, neural network performed better than moving average, linear regression and support vector regression on failure prediction\cite{iceis21}.

To help in understanding of the qualities of the task failure predictor model, the input variables of the model, including  task duration, task monetary prize, number of open tasks, and average task similarity , as well as the definition for the probability of task failure in the platform, used as the reward function to train the neural network model, are defined  below.



\textit{Def.3}: Number of Open Tasks per day ($ {NOT_{d}} $) is the Number of tasks ($ {T_{j}} $) that are open for registration when a new task ($ {T_{i}} $) arrives on the platform. 

\begin{equation}
\scriptsize
{NOT_{d}} = {\sum_{j=0}^{n}{T_{j}}}
\end{equation}
$${where: { } {TRE_{j} >= TR_{i}} }$$

\textit{Def.4}: Average Task Similarity per day ($ {ATS_{d}} $) is the average similarity score ($ {Sim_{i,j}} $) between the new arriving task ($ {T_{i}} $) and currently open tasks ($ {T_{j}} $) on the platform.  

\begin{equation}
\scriptsize
{ATS_{d}} = {\frac{\sum_{i,j=0}^{n}{Sim_{i,j}}}{NOT_{d}}}
\end{equation}
$${where: { } {TRE_{j} >= TR_{i}} } $$

\textit{Def.5}: Probability of task failure  per day ($ {p(TF_{d})} $), is the probability that a new arriving  task ($ {T_{i}} $) does not receive a valid submission and fails given its arrival date.

\begin{equation}
\scriptsize
{p(TF_{d})} =  1 - {\frac{\sum_{i=0}^{n}{VS_{i}}}{NOT_{i}}}
\end{equation}
$${where:{ } {TRE_{j} >= TR_{i}} } $$

To determine the optimal arrival date, we run the neural network model and evaluate the probability of task failure for three days (i.e two days surplus) from task arrival,  \textit{\textbf{arrival day}}: (${p(TF_{d})}$), \textit{\textbf{one day after}}:(${p(TF_{d+1})}$), and \textit{\textbf{two days after}}: (${p(TF_{d+ 2})}$). To predict the probability of task failure in future days, there is a need to determine the number of expected arriving tasks and associated task similarity scores compared to the open tasks in that day.

\textit{Def.6}: Considering that the registration duration (difference between opening and closing dates) for each task is known at any given point in time, the rate of task arrival per day ($ {TA_{d}} $) is defined as the ratio of the number of open tasks per day $ {NOT_{d}} $ over the total duration of open tasks per day.

\begin{equation}
\scriptsize
{TA_{d}} = {\frac{NOT_{d}}{\sum_{j=0}^{n}{D_{j}}}}
\end{equation}


By knowing the rate of task arrival per day, the number of open tasks for future days can be determined.

\textit{Def.7}: Number of Open Tasks in the future $ {OT_{fut}} $ is the number of tasks that are still open given a future date $ {NOT_{fut}} $, in addition to the rate of task arrival per day $ {TA_{d}} $ multiplied by the number of days into the future $ {\Delta}{days} $.

\begin{equation}
\scriptsize
{OT_{d+i}} = {{NOT_{d+i}}+{{TA_{d}}}*{\Delta}{days}}
\end{equation}

Also there is a need to know the average task similarity in future days. 

\textit{Def.8}: Average Task Similarity in the future $ {ATS_{fut}} $, is defined as the number of tasks that are still open given a future date $ {NOT_{fut}} $ multiplied by the average task similarity of this group of tasks $ {ATS_{fut}} $, the average task similarity of the current day $ {ATS_{d}} $ multiplied by the rate of task arrival per day $ {TA_{d}} $ and the the number of days into the future $ {\Delta}{days} $.

\begin{equation}
\scriptsize
{ATS_{fut}} = {{NOT_{fut}}*{ATS_{fut}}+{ATS_{d}}*{TA_{d}}*{\Delta}{day}}
\end{equation}

\subsection{Evolutionary Task Scheduling Design}

\subsubsection{Chromosome Representation and Initial Population}
The chromosome is represented as a sequence of integers where each value indicates the arrival day for the respective task. One chromosome represents the schedule for the given project. The integer values are bounded between 0 and the maximum allowed duration for the project.


\subsubsection{Reproduction and Variation}
Chromosomes are reproduced either through simple copying or through crossover, followed by mutation. We employ standard two-point crossover: two indices are selected at random and sequences between those two points are exchanged between two parent chromosomes to create two new offspring. The probability of reproduction through crossover is 0.9.
For mutation, we use a shuffling strategy. For each value in the sequence, an index is chosen randomly for shuffling. Probability for each index to get selected for shuffling is $1/(length Of Sequence)$; and, the probability of performing variation operation on a given sequence is 0.1.

During initialization of chromosomes, as well as during reproduction, we ensure that every chromosome adheres to the task dependency constraint.
The task dependency constraint is that tasks in the crowd project follow their required dependencies.

\begin{equation}
\scriptsize
{TR_{j}} = 
	\begin{cases}
        {TR_{i}} &{Sl_{i,j}} = 0 \\
        {TS_{i}} + 1  &{Sl_{i,j}} = 1
    \end{cases}
\end{equation}

\subsubsection{Fitness Functions}
We use three fitness functions that seek to, respectively, minimize project duration, minimize probability of task failure in the platform, and manage task similarity in the project. Each fitness function is described below:

\begin{itemize}

    \item \textit{Project duration}: The first fitness function measures the complete duration of the project.
    

    \item \textit{Task Similarity}: Tasks with similarity of 60\% lead to the highest level of competition with lower level of task failure \cite{10.1007/978-3-030-50017-7_7}. The similarity fitness function assures that parallel tasks that are not 60\% similar do not arrive at the same time.   

\begin{equation}
\scriptsize
	\begin{cases}
        {Pl_{i,j}} = 1 & {TR_{j}} =
            \begin{cases}
                    {TR_{i}} &{Sim_{i,j}} = 0.6 \\
                    {TRE_{i}} &{Sim_{i,j}} $$ \neq $$  0.6
            \end{cases} \\
        {Pl_{i,j}} = 0 & {TR_{j}} = {TS_{i}} + 1
        
    \end{cases}
\end{equation}
    
    \item \textit{Task Registration Start Date}: The system recommend the arrival date with the lowest probability of task failure based on the result of the neural network.   
\begin{equation}
{TR_{i}} = {min(p({TF_{d}}),}\\
{p({TF_{d+1}}),}\\
{p({TF_{d+2}}))}
\end{equation}

\end{itemize}

\subsubsection{Selection Operator}
For parent selection operation we use a crowding-distance based tournament strategy. A crowding-distance is a measure of how close a chromosome is to its neighbors. For selecting individuals for the next generation from the pool of current population and offspring we use the standard selection method from NSGA-II \cite{996017}.

\subsubsection{Termination Criterion}

In order to obtain a set of possible schedules, the system must be run multiple times. In this paper we ran the system for 200 times, at which point progress was generally observed to have stopped.

\section{Experiment Design}

To evaluate the evolutionary task scheduling method this section presents the experiment design and evaluation base line for this study.

\subsection{Research Questions}

To investigate the impact of probability of task failure in platform and task similarity level with in the project, the following research questions were formulated and studied:

\textbf{RQ1 (Overall Performance)}: \textit{How does the evolutionary task scheduling method help to reduce project duration and task failure potential?}
This research question aims to investigate the effect of the proposed method on project duration while only decreasing task failure probability in the CSD platform.  

\textbf{RQ2 (Task Similarity)}: \textit{What is the effect of introducing a task similarity objective on the algorithm's ability to find short and robust schedules?}
Understanding impact of task inter-dependencies on project duration helps to provide better strategies in task planning. 


\begin{table*}[!ht]
\scriptsize
\caption{Summary of Task Features in the Data Set} 
\centering 
\begin{tabular}{p{1.5cm} p{3cm} p{7.5cm}}
\hline
Type & Metrics & Definition \\ 
\hline
\multirow{16}{*}{\rotatebox[origin=c]{55}{Tasks Attributes}} & 
Task registration start date (TR) 	& The first day of task arrival in the platform and when workers can start registering for it. (mm/dd/yyyy) \\
& Task submission end date (TS) & Deadline by which all workers who registered for task have to submit their final results. (mm/dd/yyyy) \\

& Task registration end date (TRE)  & The last day that a task is available to be registered for.  (mm/dd/yyyy)\\
& Task Duration ($ {D_{i}} $) & total available days from task (i) registration start date ($ {TR_{i}} $) to submissions end date ($ {TS_{i}} $) \\
& Monetary Prize (MP) & Monetary prize (USD) offered for completing the task and is found in task description. Range: (0, $\infty$) \\
& Total Monetary Prize ($ {TMP_{i}} $)& The summation of the  prize that the winner ($ {MPW_{i}} $)(i.e first place) and runner up($ {MPR_{i}} $)(i.e second place) will receive after passing peer review \\
& Technology (Tech) & Required programming language to perform the task. Range: (0, \# Tech) \\ 
& Platforms (PLT) & Number of platforms used in task. Range: (0, $\infty$). \\
\hline
\multirow{8}{*}{\rotatebox[origin=c]{55}{Tasks Performance}} & Task Status & Completed or failed tasks \\
& \# Registration (R)  & Number of registrants that sign up to compete in completing a task before registration deadline. Range: (0, $\infty$). \\
& \# Submissions (S) & Number of submissions that a task receives before submission deadline. Range: (0, \# registrants]. \\
& \# Valid Submissions (VS) & Number of submissions that a task receives by its submission deadline that have passed the peer review. Range: (0, \# registrants]. \\
\hline
\label{metrics}
\end{tabular}
\end{table*}
\small

\subsection{Data Set}

The gathered data set contains 403 individual projects including 4,908 component development tasks and 8,108 workers from Jan 2014 to Feb 2015, extracted from TopCoder website. 
Tasks are uploaded as competitions in the platform and crowd software workers register and complete the challenges. On average, most of the tasks have a life cycle of 14 days from the first day of registration to the submission’s deadline. When a worker submits their final files, their submission is reviewed by experts and labeled as a valid or invalid submission.  
Table \ref{metrics} summarizes the task features available in the data set.

\subsection{Implementation of The Evolutionary Task Scheduling}

There are three steps in implementing the evolutionary task scheduler:  initial population and chromosome representation, fitness functions, and task scheduler.

\subsubsection{Initial Population and Chromosome Representation}
Task population is equal to the all the tasks in a crowdsourced project. Task execution follows the project schedule provided for a project in our data set and the initial chromosome follows the original task dependencies with 0 days delay (i.e the earliest project schedule).

\subsubsection{Fitness Function}
In order to manage fitness function in the experiment we followed below steps:

\begin{itemize}

    \item \textit{Project Duration}: The first fitness function measures the complete duration of the project for suggested schedule.

    \item \textit{Task Similarity}: The similarity function designed to calculate the cosine similarity among the tasks in the project. When two task in the project are arriving \textit{parallel} that are not following 60\% rule, task (j),$ {T_{j}} $, with longer registration duration, will be postponed to arrive after task (i), $ {T_{i}} $ registration end date.
    
   \begin{equation}
    \scriptsize
    {TR_{j}} = {TRE_{i}}
    \end{equation}
    \[where:
    \scriptsize
	\begin{cases}
        {Pl_{i,j}} = 1 \\
        {TRE_{i}} <= {TRE_{j}}
    \end{cases}
    \]
    
    \item \textit{Task Failure Probability}: Tasks following the dependency requirement function and meeting the task similarity are entering the task failure probability function to be assigned to the date with \textit{lowest} failure probability.

\end{itemize}

\begin{figure*}[ht!]
\scriptsize
\centering
\includegraphics[width=1\textwidth,height=0.5\textwidth]{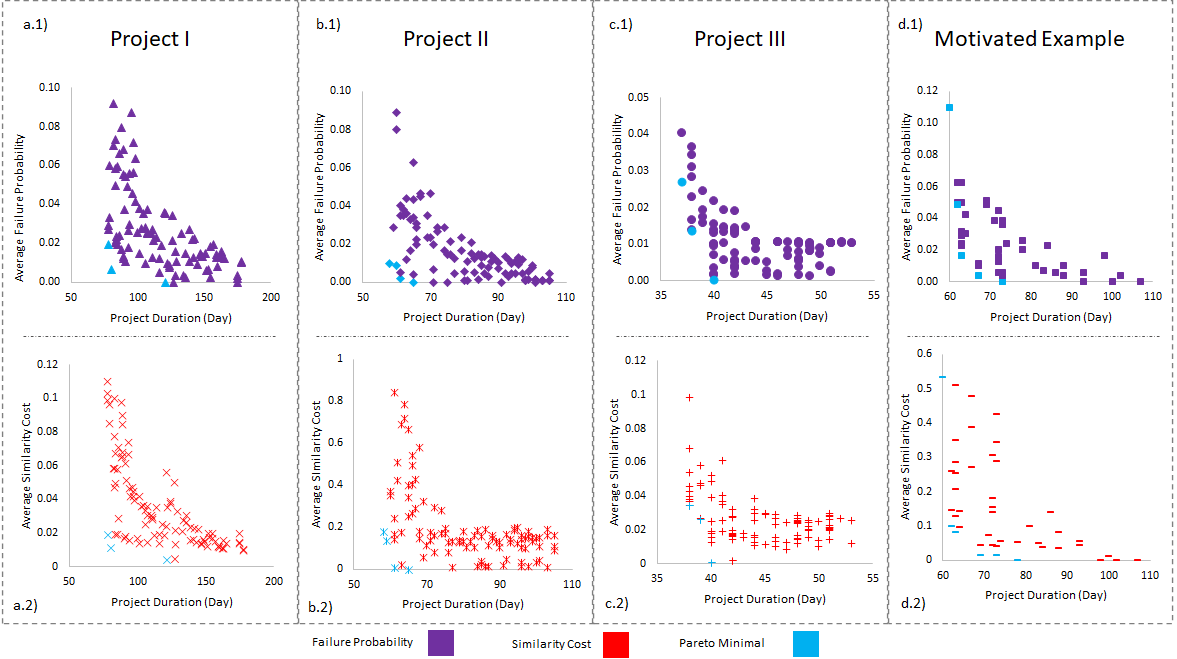}
\caption{Average Task failure Probability (a.1, b.1, c.1, d.1) and Task Similarity Cost(a.2, b.2, c.2, d.2) of Recommended Schedules per project}
\label{paret}
\end{figure*}

\section{Results}
This section presents the evaluation results of applying the proposed scheduling framework to 4 CSD projects. Table \ref{projects} summarize the projects overall view. As figure \ref{paret}-a. represents, the framework recommend a schedule with duration of 121 days with 0.0\% probability of task failure and 0.4\% similarity cost for project I. This provides almost 70\% schedule acceleration. The result of scheduling for project II, figure \ref{paret}-b., lead to a project duration with pareto minimum of 65 day, and 0\% failure probability and task similarity cost. 

The recommended schedule results in 78\% acceleration. Applying the presented framework to the project III, \ref{paret}-c. lead to 55\% shorter schedule acceleration. the project duration shorten to 40 days, with probability of task failure as low as 0.02\% and task similarity cost of 0.04\%. Finally, scheduling the motivating example with the presented system recommended a schedule with 73 days duration, 0.07\% average probability of task failure and 1.3\% average task similarity cost.The recommended schedule leads to 33\% schedule acceleration.

\begin{small}
    \begin{table}[!ht]
    \scriptsize
    \caption{Summary of Scheduled Projects} 
    \centering 
    \begin{tabular}{p{1.5cm} p{1.5cm} p{1.5cm} p{1.5cm} p{1.5cm} p{1.5cm}}
    \hline
   Project ID  & \# Tasks &  Original (day) & Final (day) & Recom (day) & Schedule Acceleration\\ 
    \hline
    Project I & 25 & 70 & 393 & 121 & 70\% \\
    Project II & 24 & 58 & 203 & 65 & 78\% \\
    Project III & 23 & 31 & 88 & 40 & 55\%\\
    Motivating Example & 11 & 37 & 110 & 73 & 33\% \\
  
    \hline
    \label{projects}
    \end{tabular}
    \end{table}
\end{small}


In order to  answer the research questions in part V-A and have a better understanding of how the presented framework part IV-A is working, we explain  scheduling of motivating example using the evolutionary scheduling introduced in part IV-A in details. 


\begin{figure}[ht!]
\centering
\includegraphics[width=1\textwidth]{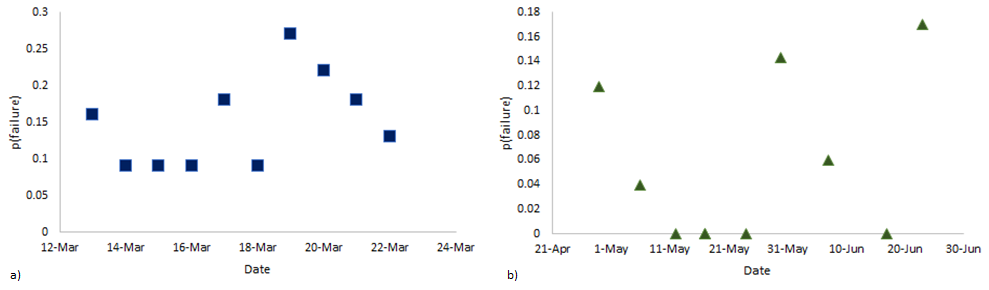}
\caption{a) Overall Performance of scheduling framework, b) Schedule Acceleration with Task Similarity}
\label{SA}
\end{figure}

\subsection{Project Schedule Acceleration}
To answer RQ1 we eliminate the second fitness function (\textit{task similarity}) and scheduled the project focusing on minimizing task failure probability.
As it is shown in figure \ref{SA}-a, project duration has decreased from 110 days to 60 days.The probability of task failure has dropped to 10\%. While the scheduling method providing 23 days delay in compare with the shortest possible project plan(37 days), the recommended schedule finishes 50 days earlier than the original project duration with 47\% higher chance of success. Moreover, according to the  recommended project timeline, project faced the maximum task failure probability of 33\% on task 7  and the minimum failure probability of 0.0 on tasks 2,5,6,8,9 and 10.

\subsection{Task Similarity}
To answer RQ2 we have added task similarity fitness function to the evolutionary scheduling. In this part we analyze the recommended schedule with 0 similarity cost form the solution space.
As it is presented in figure \ref{SA}-b, evolutionary scheduling recommends a plan with duration of 79 days and average probability of task failure of 5\%. Compare with the schedule from RQ1, evolutionary scheduling provides 19 days longer project plan, while it provides 5\% lower task failure probability. Also, evolutionary scheduling takes similarity among the project tasks in to account.This creates easier competition for attracting resources.  
Also, the method recommended to postpone the start date of the project  for 47 day,
in order to have 0 days overlap due to task similarity.


\subsection{Comparison between Scheduling with and without Task Similarity}
To have a a better understanding of the impact of the proposed evolutionary scheduling we investigate the number of open tasks, number of open  tasks and average task similarity level on the recommended schedule. 

\begin{figure}[ht]
\centering
\scriptsize
\includegraphics[width=1\textwidth,keepaspectratio]{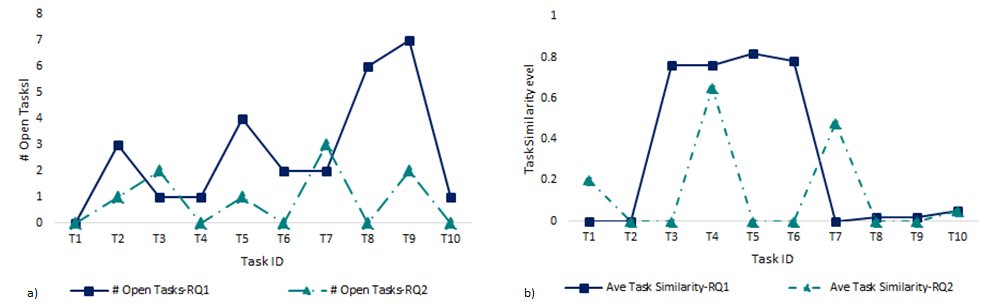}
\caption{a) Number of Open Tasks on Task Arrival Day, b) Average Task Similarity on Task Arrival Day}
\label{OT}
\end{figure}

Figure \ref{OT}-a presents the number of open task per arrival task based on different scheduling plan in RQ1 and RQ2. Scheduling under RQ1 conditions
lead to competition level with on average 3 other tasks per day. The least competition to attract the resources happened for $ {T_{1}} $ with 0 open tasks in the arrival date and $ {T_{9}} $ faced the hardest competition with 7 other tasks available in the platform.
Clearly, taking task similarity into account creates less competition. 
As it is shown scheduling under RQ2 provides task arriving with average one open task per arrival date, with minimum 0 open task for $ {T_{1}} $,$ {T_{4}} $,$ {T_{6}} $,$ {T_{8}} $,$ {T_{10}} $, and maximum 3 open tasks when $ {T_{7}} $ arrived.

Another measure to investigate is average task similarity on the recommended task arrival date. As it is shown in figure \ref{OT}-b, in RQ1
tasks are arriving to the platform with on average 32\% task similarity. $ {T_{3}} $,$ {T_{4}} $,$ {T_{5}} $, and$ {T_{6}} $ are competing with tasks with more than 75\% similarity. While adding task similarity fitness function in RQ2 provides task arrival with on average 14\% task similarity in the platform.$ {T_{4}} $ faced the highest task similarity level of 65\%.



\subsection{Discussion and Findings} 
Our evolutionary scheduling method generated plans with average probability of task failure of 5\%. This result not only is 53\% lower than the original plans, but it is also lower than reported failure ratio in the platform (i.e 15.7\%) \cite{mejorado2020study}\cite{yang2016should}.
Figure \ref{paret}-d-1. shows the average probability of task failure per recommended schedule by evolutionary scheduler. According to the figure \ref{paret}-d-1 longer project duration provides lower probability of task failure. The goal is to reducing schedule acceleration while reducing task failure, hence, the optimum solutions occurs in under recommended schedule with duration around 73 days (33\% schedule acceleration) which provides task failure probability of 0.07\%(Pareto minimal solutions shown in blue) .    


Investigating the number of open tasks and the average task similarity on recommended arriving day per task support that the evolutionary task scheduling method assures lower competition over shared supplier resources per arrival task(i.e 1 and 14\% respectively). 
Task similarity cost is the ratio of duration added to the recommended task schedule due to considering task similarity fitness function. As it is shown in figure \ref{paret}-d-2 the lowest similarity cost happens when either project duration is between 70 days to 80 days or more than 95 days. Since the main objective is reducing project during the optimum solution is project duration of 70-80 days(pareto minimal shown in blue).


\subsection{Threats to Validity}

\textit{First}, the study only focuses on competitive CSD tasks on the TopCoder platform. Many more platforms do exist and there is no guarantee the same results would remain exactly the same in other CSD platforms.
\text{Second}, there are many different factors that may influence task similarity, task success, and task completion. Our similarity algorithm and task failure probability-focused approach are based on known task attributes in TopCoder. Different similarity algorithms and task failure probability-focused approaches may lead us to different, but similar results.
\textit{Third}, the result is based on tasks only. Workers' network and communication capabilities are not considered in this research. 

\subsection{Adaptability to Different Platforms}
The overall presented scheduling framework in this paper is adaptable to different crowdsourced platforms. However based on the type of platform 
the task failure fitness needs to be updated by 
the chosen platform.
To make the presented framework compatible for different crowdsourcing platforms, there is no need to update any part of the scheduling framework but the back end analysis in elaborating defined input metrics.

\section{Related Work}

Delay scheduling \cite{tajedin2013determinants} is the first presented scheduling method for crowdsourced projects to maximize the probability of a worker receiving tasks from the same batch of tasks they were working on. An extension of this idea is fair sharing schedule \cite{alba2006efficient}. In that method, heterogeneous resources would be shared among all tasks with different demands, which ensures that all tasks would receive the same amount of resources to be fair. 
Later on, weighted fair sharing (WFS) \cite{barreto2008staffing} was presented as a method to schedule batches based on their priority. Tasks with higher priority are introduced first. Another proposed crowd scheduling method is quality of submissions (QOS) \cite{khazankin2011qos}, a skill-based scheduling method with the purpose of minimizing scheduling while maximizing quality by assigning the task to the most available qualified worker. This method was created by extending standards of Web Service Level Agreement (WSLA) \cite{regnell2005market}. The third available method is game with a purpose \cite{cusumano2004business}, in which a task will not be started unless a certain number of workers registered for it.
Similarly, CrowdControl \cite{rajan2013crowdcontrol}, provides a scheduling approach that assigns
tasks to workers not only based on their historical performance but also their learning curve of performing tasks. SmartCrowd \cite{roy2014optimization} schedules tasks based on worker skills and their reward requirements to perform the task. Using above scheduling methods, HIT-Bundle (Human Intelligent Task) \cite{difallah2016scheduling} provides a batch container which schedules heterogeneous tasks into the platform from different batches. This method makes a more successful outcome by applying different scheduling strategies at the same time. The most recent method is helping crowdsourcing-based service providers to meet completion time SLAs \cite{hirth2019task}. That system works based on the oldest task waiting time and runs a stimulative evaluation to recommend best scheduling strategy in order to reduce the task failure ratio.

\section{Conclusion and Future Work}

CSD provides software organizations access to an infinite, online worker resource supply. Assigning tasks to a pool of unknown workers from all over the globe is challenging. A traditional approach to solving this challenge is task scheduling. Improper task scheduling in CSD may cause Task failure due to uncertain worker behavior.
The proposed approach recommends task scheduling plans based on a set of task dependencies in a crowdsourced project, similarities among tasks, and task failure probabilities based on recommended arrival date. The proposed evolutionary scheduling method utilizes a genetic algorithm to optimize and recommend the task schedule. The method uses three fitness functions, respectively based project duration, task similarity, and task failure prediction. The task failure fitness uses a neural network to predict probability of task failure on arrival date. 
The proposed method empowers crowdsourcing managers to explore potential outcomes with respect to different task arrival strategies. This includes the probability of task failure, number of open similar task in terms of context, prize, duration and type on the arrival day, and different schedule acceleration.The experimental results on 4 different projects demonstrate that the proposed method reduced project duration on average 59\% 

In future, we would like to expand our model to provide a platform level scheduler which not only recommends the best scheduling plan but also provides required task decomposition action to achieve the lowest task failure.

 \bibliography{referencefile.bib}

\end{document}